# Unraveling the Angular Symmetry of Optical Force in a Solid Dielectric

Xiang Xi,[1] Jingwen Ma,[1] Zhong-Hao Zhou,[2,3] Xin-Xin Hu,[2,3] Yuan Chen,[2,3] Chang-Ling Zou,[2,3,*] Chun-Hua Dong,[2,3,*] and Xiankai Sun[1,*]

[1]Department of Electronic Engineering, The Chinese University of Hong Kong, Shatin, New Territories, Hong Kong

[2]CAS Key Laboratory of Quantum Information, University of Science and Technology of China, Hefei, Anhui 230026, China

[3]CAS Center for Excellence in Quantum Information and Quantum Physics, University of Science and Technology of China, Hefei, Anhui 230026, China

[*]Corresponding author. Email: clzou321@ustc.edu.cn (C.L.Z.); chunhua@ustc.edu.cn (C.H.D.); xksun@cuhk.edu.hk (X.S.)

**The textbook-accepted formulation of electromagnetic force was proposed by Lorentz in the 19th century, but its validity has been challenged due to incompatibility with the special relativity and momentum conservation. The Einstein–Laub formulation, which can reconcile those conflicts, was suggested as an alternative to the Lorentz formulation. However, intense debates on the exact force are still going on due to lack of experimental evidence. Here, we report the first experimental investigation of angular symmetry of optical force inside a solid dielectric, aiming to distinguish the two formulations. The experiments surprisingly show that the optical force exerted by a Gaussian beam has components with the angular mode number of both 2 and 0, which cannot be explained solely by the Lorentz or the Einstein–Laub formulation. Instead, we found a modified Helmholtz theory by combining the Lorentz force with additional electrostrictive force could explain our experimental results. Our results represent a fundamental leap forward in determining the correct force formulation, and will update the working principles of many applications involving electromagnetic forces.**

The force exerted by electromagnetic fields is of fundamental importance in broad sciences and applications [1-3], but its exact formulation inside media is still controversial and unclear [4-6]. The Lorentz (LO) law of electromagnetic force is widely adopted and regarded as one of the foundations of classical electrodynamics. However, this century-old physical law has been in crisis [7]. In the 1960s, Shockley pointed out that the LO law contradicts the universal momentum conservation in certain systems involving magnetic media [8-10]. More recently, the LO law was also found to be incompatible with the special relativity, as it predicts different results in different reference frames [11]. These problems of the LO law could be avoided by introducing an additional hidden momentum of electromagnetic field in magnetic media [8,11]. However, there still lack wide agreements on this issue because the hidden momentum is experimentally unobservable with current technique. At the same time, another formulation originally proposed by Einstein and Laub (EL) has also been widely used and was suggested as an alternative of the electromagnetic force formulation [4,11-21], as it complies with both the special relativity and universal conservation laws without needing the hidden momentum [11,22,23]. The EL formulation is also consistent with the Maxwell's equations, and agrees with the existing measurement results of the total force or torque that support the LO formulation [19,24]. Their equivalence on the total force or torque measurements leads to most of the existing experiments [4-6] failing to distinguish these two formulations. To date, the debates on the LO and EL formulations are still going on because rigorous experimental investigations on distinguishing them are still absent.

The underlying difference between the LO and EL formulations lies in their different descriptions of the quantum nature of media and electromagnetic fields: the LO formulation treats the electric and magnetic dipoles inside a medium as distributions of ordinary charges and currents, while the EL formulation treats the electric and magnetic dipoles as two individual constituents that are distinct from ordinary charges and currents [17,19]. Due to the different treatments, the LO force in a nonmagnetic dielectric material has the form $\mathbf{F}_{\mathrm{LO}} = (-\nabla \cdot \mathbf{P})\mathbf{E} + \partial \mathbf{P}/\partial t \times \mathbf{B}$, while the EL force has the form $\mathbf{F}_{\mathrm{EL}} = (\mathbf{P} \cdot \nabla)\mathbf{E} + \partial \mathbf{P}/\partial t \times \mathbf{B}$, with $\mathbf{E}$ the electric field, $\mathbf{B}$ the magnetic induction, and $\mathbf{P} = \varepsilon_0(\varepsilon_{\mathrm{r}} - 1)\mathbf{E}$ the polarization (Supplementary Sec. 2). $\varepsilon_0$ and $\varepsilon_{\mathrm{r}}$ are the vacuum permittivity and the relative permittivity of the material, respectively. Note that the hidden momentum problem can be avoided naturally in nonmagnetic dielectric media, inside which the hidden momentum is always zero. It was recently discovered that although these two formulations predict the same total force on an object, they actually produce different force distributions inside a dielectric medium [19,21]. This feature can be harnessed in experiments to distinguish the two formulations. However, the predicted differences are microscopic and exist only inside a medium, which were thought to be too weak to be detected.

Here, we investigated for the first time the optical force distribution inside a solid dielectric by employing an optomechanical approach with ultrahigh detection sensitivity. Theoretically, the optical force distribution exerted by a linearly polarized optical Gaussian beam inside a dielectric has angular symmetry with angular mode number $C = 2$ by the LO formulation or $C = 0$ by the EL formulation (Fig. 1). We derived three criteria for determining the angular symmetry of optical force distribution inside a single-mode optical fiber. Surprisingly, multiple experiments based on these three criteria all show that the optical force distribution of a Gaussian beam in an optical fiber has components of both $C = 2$ and $C = 0$. These results cannot be explained solely by the LO or the EL formulation, indicating the necessity of a modification or a new theory. We found that a modified Helmholtz theory by supplementing the LO force with additional electrostrictive force may explain the experimental results. Our experiment in a solid dielectric represent a fundamental leap forward in experimental exploration of the optical force distribution, because it can avoid many spurious effects in previous experiments [4,21], can have ultrahigh sensitivity, and can identify different optical force components separately. Our results will not only play an important role in determining the correct formulation of electromagnetic force, but also provide a scheme to solve some other issues in classical electrodynamics, such as the Abraham–Minkowski controversy.

**Angular mode number of the optical force**. For a linearly polarized optical beam of Gaussian profile propagating in a dielectric medium, such as a single-mode fiber [Fig. 1(a)], the LO formulation predicts a force density distribution tending to stretch (compress) the medium along (perpendicular to) the light polarization direction [Fig. 1(b)], and the EL formulation predicts a force density distribution tending to compress the medium radially inward [Fig. 1(c)] [19,21]. Such force density distribution in the LO formulation has a form in the cylindrical coordinates ($\mathbf{r}$: $r$, $\theta$, $z$) as

$$\mathbf{F}_{\mathrm{LO}}(\mathbf{r}) = \begin{pmatrix} f_r^{\mathrm{LO}}(r)\cos 2(\theta + \phi) \\ -f_r^{\mathrm{LO}}(r)\sin 2(\theta + \phi) \\ 0 \end{pmatrix},$$

(1)

and the force density in the EL formulation is

$$\mathbf{F}_{EL}(\mathbf{r}) = \begin{pmatrix} f_r^{EL}(r) \\ 0 \\ 0 \end{pmatrix}, \qquad (2)$$

with $\phi$ the polarization angle of the optical beam (Supplementary Sec. 2). The functions $f_r^{LO}(r)$ and $f_r^{EL}(r)$ are related to the optical fields and vary only along the radial direction. According to azimuthal properties described in Eqs. (1) and (2), the optical force density mode possesses angular symmetry with the angular mode number $C = 2$ by the LO formulation [Fig. 1(b)] or $C = 0$ by the EL formulation [Fig. 1(c)]. Therefore, one can distinguish these two formulations by experimentally measuring the angular symmetry of the optical force instead of the absolute mechanical displacement. The absolute mechanical displacement is intrinsically extremely weak and can easily be masked by noises, but the angular mode number of the force is robust and can unambiguously be determined as an integer.

We employed an optical-fiber-based system to identify the angular symmetry of the optical force in a slightly modified single-mode fiber [Fig. 1(a)]. In the system, the optical force was exerted by linearly polarized optical fields propagating in the core of the fiber. The intensity ($E_0^2$) of the optical field was sinusoidally modulated (with frequency $\Omega$, modulation depth $A$, and RF modulation phase $\varphi_{RF}$) to generate oscillating optical force to actuate the mechanical modes [Fig. 1(d)] of the fiber. The oscillating part of the optical force can be described as $\mathbf{F}(\mathbf{r}, t) = A E_0^2 \cos(\Omega t + \varphi_{RF}) \cdot \mathbf{F}_{LO/EL}(\mathbf{r})$ Due to the resonant enhancement effect, the mechanical modes could have amplified mechanical motion in response to the force oscillating at the mechanical eigenfrequencies. The intensities of the actuated mechanical modes were obtained with ultrahigh sensitivity from optomechanical transduction by using an ultrahigh-$Q$ optical whispering-gallery mode traveling in the circumference of the transverse plane, which was supported by the slightly fused cladding of the optical fiber (Fig. 1a) [25].

According to Eqs. (1) and (2), a critical property of optical force is that the LO force with $C = 2$ is dependent on the optical polarization angle $\phi$ while the EL force with $C = 0$ is independent of $\phi$. Therefore, the mechanical modes actuated by the optical force would have different predicted response to optical polarization angle $\phi$ for these two theories. Solving the elastic equation, it is found the actuated amplitude $x_{amp}$ of a mechanical mode is proportional to the spatial overlap integral of the force density distribution and mechanical modal profile (Supplementary Sec. 3.1):

$$x_{amp} \propto \iiint \mathbf{F}_{EL/LO}(\mathbf{r}) \cdot \mathbf{u}(\mathbf{r}) d\mathbf{r}, \tag{3}$$

where $\mathbf{u}(\mathbf{r})$ is the displacement modal profile of the mechanical mode. Here, we focus on the response of the mechanical wine-glass mode and breathing mode of the fiber (Fig. 1d). The mechanical displacement modal profile of these mechanical eigenmodes can be expressed as [26]

$$\mathbf{u}(\mathbf{r}) = \begin{pmatrix} \left[U_m(r) + \Delta U\right]\left[\cos n(\theta + \phi_m) + \sum_l \sigma_l^{(n)} \cos l(\theta + \phi_m)\right] \\ \left[V_m(r) + \Delta V\right]\left[\sin n(\theta + \phi_m) + \sum_l \sigma_l^{(n)} \sin l(\theta + \phi_m)\right] \\ \left[W_m(r) + \Delta W\right]\left[\cos n(\theta + \phi_m) + \sum_l \sigma_l^{(n)} \cos l(\theta + \phi_m)\right] \end{pmatrix} e^{-|z|/\Gamma_z}, \tag{4}$$

where $U_m(r)$, $V_m(r)$, and $W_m(r)$ have complicated expressions involving the Bessel functions, $\phi_m$ is the angle of symmetry axis of the mechanical mode and can be set as 0° for convenience, $\Gamma_z$ is the evanescent length of elastic wave in the $z$ direction (Supplementary Sec. 1.4). $\Delta U$, $\Delta V$, $\Delta W$, $\sum_l \sigma_l^{(n)} \cos l(\theta + \phi_m)$, and $\sum_l \sigma_l^{(n)} \sin l(\theta + \phi_m)$ all represent the perturbed modal distortion due to the small geometric imperfection of the device. Note these perturbation terms are added for a complete description but are not necessary. The mechanical wine-glass mode has an angular mode number $n = 2$ and breathing mode has $n = 0$, whose displacement fields have similar angular symmetry to that of the force density in LO and EL formulations [Figs. 1(b)–(d)], respectively. Such resemblance of angular symmetry actually determines the spatial overlap integral in Eq. (3) —specifically, the mechanical mode with angular mode number $n$ can only respond selectively to the force with $C = n$ in ideal case. Therefore, the wine-glass mode with $n = 2$ is employed to determine whether there exists component of force density with angular mode number $C = 2$, and the breathing mode with $n = 0$ is employed to determine the existence of component of force density with $C = 0$. Based on mechanical response determined by the integral in Eq. (3), we summarized three criteria for determining the angular mode number $C$ of the optical force density inside the fiber (Supplementary Sec. 3):

(I) For a single pump beam with polarization angle $\phi$, the intensity of mechanical mode actuated by a force with $C = 2$ is proportional to $\left|\cos(2\phi)\right|^2$, while that by a force with $C = 0$ is polarization-independent.

(II) For dual pump beams with polarization angles $\phi_1$ and $\phi_2$ and the same modulation phase $\varphi_{\mathrm{RF}}$, the intensity of mechanical mode actuated by the two forces with $C = 2$ is proportional to $\left|\cos(2\phi_1) + \cos(2\phi_2)\right|^2 = \left|2\cos(\phi_1 + \phi_2)\cos(\phi_1 - \phi_2)\right|^2$, while that by forces with $C = 0$ is polarization-independent.

(III) For dual orthogonally polarized pump beams with a RF modulation phase difference $\Delta\varphi_{\mathrm{RF}} = \varphi_{\mathrm{RF2}} - \varphi_{\mathrm{RF1}}$, the superimposed force is $AE_0^2[\cos(\Omega t + \varphi_{\mathrm{RF2}}) - \cos(\Omega t + \varphi_{\mathrm{RF1}})] \cdot \mathbf{F}_{\mathrm{LO}}(\mathbf{r})$ or $AE_0^2[\cos(\Omega t + \varphi_{\mathrm{RF2}}) + \cos(\Omega t + \varphi_{\mathrm{RF1}})] \cdot \mathbf{F}_{\mathrm{EL}}(\mathbf{r})$ because of the relation $\mathbf{F}_{\mathrm{LO},\phi} = -\mathbf{F}_{\mathrm{LO},\phi + \pi/2}$ and $\mathbf{F}_{\mathrm{EL},\phi} = \mathbf{F}_{\mathrm{EL},\phi + \pi/2}$. The amplitude of the superimposed force is then proportional to $\sin(\Delta\varphi_{\mathrm{RF}}/2)$ for LO force with $C = 2$ and $\cos(\Delta\varphi_{\mathrm{RF}}/2)$ for EL force with $C = 0$. Therefore, the intensity of mechanical mode actuated by such two forces with $C = 2$ would be proportional to $|\sin(\Delta\varphi_{\mathrm{RF}}/2)|^2$, while that by forces with $C = 0$ is proportional to $|\cos(\Delta\varphi_{\mathrm{RF}}/2)|^2$.

**Measurement of the angular symmetry of optical force**. To experimentally examine the angular mode number of the optical force under the three criteria, we fabricated a bottle-like microstructure on a standard single-mode optical fiber (Fig. 2; Supplementary Sec. 1). We slightly fused the cladding of the fiber to create the two necks of the bottle-like microstructure [Fig. 2(a)], whose diameters range from 100 to 120 μm for different samples tested in the experiment (Supplementary Fig. S1). Such bottle-like device configuration forms optical and mechanical energy potentials for supporting high-quality optical probe modes and mechanical modes (Supplementary Secs. 1.3 and 4.3). Obtaining high quality factors in these optical probe modes and mechanical modes is crucial for achieving high sensitivity in detecting the mechanical motion, because the optomechanical transduction and resonant amplification of the mechanical motion depend respectively on the optical and mechanical quality factors. The pump light beam that exerts an optical force to actuate the mechanical modes propagates in the fiber core [Fig. 2(b)], with the beam shape preserving a quasi-Gaussian profile after fabricating the bottle-like microstructure [Figs. 2(c) and S2]. The pump light is also experimentally confirmed to be quasi-linearly polarized (Supplementary Sec.

4.4). Such quasi-linearly polarized Gaussian pump light beam well satisfies all the critical experimental requirements for the theoretical analysis about force density distribution and the three criteria to examine the force formulations above (Supplementary Secs. 3 and 6).

The angular mode number of the optical force density was experimentally investigated by measuring the intensity of the wine-glass mode ($n = 2$) according to Criteria I and II. First, we measured the response of mechanical intensity to the polarization angle of a single pump beam. It was found that the mechanical intensity follows the pump beam's polarization angle $\phi$ with a dependence of $\left| \cos(2\phi) \right|^2$, with >20 dB extinction ratio [Fig. 3(a)]. Next, we applied two pump beams and measured the response of the same mechanical mode to the two pump beams' polarization angles $\phi_1$ and $\phi_2$. It was found that the mechanical intensity follows $\left| \cos(\phi_1 + \phi_2) \cos(\phi_1 - \phi_2) \right|^2$ [Fig. 3(b)], with >20 dB extinction ratio. When $\phi_2$ is fixed at 0°, the measured mechanical intensity follows a dependence of $\left| \cos\phi_1 \right|^4$ [Fig. 3(c)]. Specifically, for two orthogonally polarized pump beams ($\phi_1 = 90^\circ$, $\phi_2 = 0^\circ$), the measured mechanical intensity is much weaker than that actuated by a single pump beam ($\phi = 0^\circ$ or $90^\circ$), indicating that the forces of two orthogonally polarized pump beams cancel each other out [Figs. 3(d)–3(f)]. According to Criteria I and II, these results indicate the existence of force component with $C = 2$.

To further investigate the angular mode number of optical force, we also measured the actuation results of the breathing mode ($n = 0$) with the same experimental configuration. With a single pump beam, the mechanical intensity does not vary with the polarization angle [Fig. 4(a)]. In addition, the mechanical intensity also remains constant under actuation by dual pump beams with different polarization angles [Fig. 4(b)]. According to Criteria I and II, these results indicate that the optical force also has a component with $C = 0$.

Next, the angular mode number of optical force was also investigated under the condition in Criterion III, where the wine-glass mode ($n = 2$) and the breathing mode ($n = 0$) each were actuated by two orthogonally polarized pump beams modulated at the same RF frequency but with a constant phase difference $\Delta\varphi_{RF}$. Figure 5(a) shows the measured mechanical intensity of the wine-glass mode ($n = 2$) as a function of $\Delta\varphi_{RF}$, which follows the dependence of $\left| \sin(\Delta\varphi_{RF}/2) \right|^2$ and confirms the existence of force component with $C = 2$. On the other hand, the mechanical intensity of the breathing mode ($n = 0$) follows $\Delta\varphi_{RF}$ with a dependence of $|\cos(\Delta\varphi_{RF}/2)|^2$ [Fig. 5(b)], which confirms the existence of force component with $C = 0$.

The above experimental results indicate that the optical force by a linearly polarized Gaussian beam in a solid dielectric medium has components with angular mode number of both $C = 2$ and $C = 0$. Such results are highly reproducible, and are confirmed to be valid even when the optical and mechanical modes are imperfect due to some moderate distortion of the fabricated device structure (Supplementary Secs. 5 and 6). This is because the polarization dependence of the optical force density in the LO or EL formulation is not affected by slight geometric perturbation of the device structure. Nonetheless, a small portion of crosstalk exists due to the perturbation terms caused by the geometric imperfection, yielding actuation of the mechanical modes with $n \neq C$. By taking this factor into account, we numerically simulated the actuated mechanical intensities and compared them with the measured results, concluding that the ratio between the force components with $C = 2$ and $C = 0$ is between 1:3 and 1:1 (Supplementary Sec. 7). Therefore, these two force components are comparable in magnitude. Since the LO and EL formulations each predict an optical force with a unique angular mode number ($C = 2$ or $C = 0$), neither of them can explain our experimental results.

**Discussion**

Although the unraveled angular symmetry of optical force contradicts the predictions of both the LO and EL formulations, our results are consistent with previous experimental observation by Ashkin and Dziedzic in 1973 [27]—a bulge appeared on water surface at the spot where a focused laser beam entered, which was ever taken as an evidence supporting the EL formulation [20,21]. According to our experimental results, such a bulge can be generated as long as the angularly symmetric compressive force component with $C = 0$ exists. It should also be noted that the Hakim–Higham experiment in 1962 [28] was believed to support the Helmholtz force over that by Einstein and Laub. Actually, the Hakim–Higham experiment only showed the directionless strength of electric pressure along the $y$ axis in their setup. Such a one-dimensional scalar measurement is not enough to determine the distribution and angular mode number of electromagnetic force. Our findings can also be compatible with their results.

The coexistence of angular mode number $C = 2$ and $C = 0$ of the optical force density inside a dielectric has not been experimentally identified before, because most relevant experiments are done in liquids [4,5,29,30]. The fluidic nature of liquids makes them challenging to measure the angularly antisymmetric force component with $C = 2$, and also make them unable to provide detailed microscopic information about the force distribution. Additionally, those conventional experiments based on liquids are mostly phenomenological with some spurious effects [4,21]. By contrast, our experiment based on a lossless solid dielectric avoids most of the ambiguous effects encountered previously, and the mechanical modes of the device enable the first unraveling of the detailed microscopic properties of optical force inside a medium. We expect that these results will not only generate long-term impact on understanding of the light–matter interactions, but also update the fundamental working principle for many applications in science and engineering branches involving optical forces.

Although the experiments were planned based on the force distributions inside a medium predicted by the LO and EL formulations, the unraveled angular symmetry can be used to examine any other related theories [4-6] besides the LO and EL formulations. The force density distribution of a Gaussian beam in an optical fiber predicted by these existing theories can also have an angular mode number $C = 2$ or $C = 0$. Exhaustive scrutiny of all the force formulations, however, is beyond the scope of this work. Here, we found that a modified Helmholtz theory by combining the Lorentz formulation with the electrostrictive force [31,32] could account for the coexistence of force components with $C = 2$ and $C = 0$, which possibly explains our experimental results (Supplementary Sec. 8). On the other hand, since the EL formulation has already included the electrostrictive interaction [4,24], it may require other types of modification to explain the experimental results. We believe that the angular symmetry of the optical force unraveled in this work will serve as a crucial step in the ultimate determination of the correct electromagnetic force formulation inside media in the future.

**References**

[1]   A. Ashkin, *Acceleration and trapping of particles by radiation pressure*, Phys. Rev. Lett. **24**, 156 (1970).
[2]   S. Chu, L. Hollberg, J. E. Bjorkholm, A. Cable, and A. Ashkin, *Three-dimensional viscous confinement and cooling of atoms by resonance radiation pressure*, Phys. Rev. Lett. **55**, 48 (1985).
[3]   K. Dholakia and T. Čižmár, *Shaping the future of manipulation*, Nat. Photonics **5**, 335 (2011).

[4] I. Brevik, *Experiments in phenomenological electrodynamics and the electromagnetic energy-momentum tensor*, Phys. Rep. **52**, 133 (1979).

[5] I. Brevik, *Radiation forces and the Abraham–Minkowski problem*, Mod. Phys. Lett. A **33**, 1830006 (2018).

[6] P. W. Milonni and R. W. Boyd, *Momentum of light in a dielectric medium*, Adv. Opt. Photon. **2**, 519 (2010).

[7] A. Cho, *Textbook electrodynamics may contradict relativity*, Science **336**, 404 (2012).

[8] W. Shockley and R. P. James, *"Try simplest cases" discovery of "hidden momentum" forces on "magnetic currents"*, Phys. Rev. Lett. **18**, 876 (1967).

[9] W. Shockley, *"Hidden linear momentum" related to the $\vec{\alpha} \cdot \vec{E}$ term for a Dirac-electron wave packet in an electric field*, Phys. Rev. Lett. **20**, 343 (1968).

[10] S. Coleman and J. H. Van Vleck, *Origin of "hidden momentum forces" on magnets*, Phys. Rev. **171**, 1370 (1968).

[11] M. Mansuripur, *Trouble with the Lorentz law of force: incompatibility with special relativity and momentum conservation*, Phys. Rev. Lett. **108**, 193901 (2012).

[12] A. Einstein and J. Laub, *About the ponderomotor forces exerted on resting bodies in the electromagnetic field*, Annalen der Physik **331**, 541 (1908).

[13] A. Einstein and J. Laub, *About the basic electromagnetic equations for moving bodies*, Annalen der Physik **331**, 532 (1908).

[14] R. M. Fano, L. J. Chu, and R. B. Adler, *Electromagnetic Fields, Energy, and Forces* (Wiley, New York, 1960).

[15] J. P. Gordon, *Radiation forces and momenta in dielectric media*, Phys. Rev. A **8**, 14 (1973).

[16] G. B. Walker and D. G. Lahoz, *Experimental observation of Abraham force in a dielectric*, Nature **253**, 339 (1975).

[17] L. Vaidman, *Torque and force on a magnetic dipole*, Am. J. Phys. **58**, 978 (1990).

[18] R. Loudon, *Theory of the forces exerted by Laguerre-Gaussian light beams on dielectrics*, Phys. Rev. A **68**, 013806 (2003).

[19] S. M. Barnett and L. Rodney, *On the electromagnetic force on a dielectric medium*, J. Phys. B **39**, S671 (2006).

[20] R. Loudon, *Radiation pressure and momentum in dielectrics*, Fortschr. Phys. **52**, 1134 (2004).

[21] M. Mansuripur, A. R. Zakharian, and E. M. Wright, *Electromagnetic-force distribution inside matter*, Phys. Rev. A **88**, 023826 (2013).

[22] M. Mansuripur, *Electromagnetic force and torque in ponderable media*, Opt. Express **16**, 14821 (2008).

[23] M. Mansuripur, *Radiation pressure and the linear momentum of the electromagnetic field in magnetic media*, Opt. Express **15**, 13502 (2007).

[24] M. Mansuripur, *Force, torque, linear momentum, and angular momentum in classical electrodynamics*, Appl. Phys. A **123**, 653 (2017).

[25] M. Asano, Y. Takeuchi, W. Chen, Ş. K. Özdemir, R. Ikuta, N. Imoto, L. Yang, and T. Yamamoto, *Observation of optomechanical coupling in a microbottle resonator*, Laser Photon. Rev. **10**, 603 (2016).

[26] J. Zemanek, *An experimental and theoretical investigation of elastic wave propagation in a cylinder*, J. Acoust. Soc. Am. **51**, 265 (1972).

[27] A. Ashkin and J. M. Dziedzic, *Radiation pressure on a free liquid surface*, Phys. Rev. Lett. **30**, 139 (1973).

[28] S. S. Hakim and J. B. Higham, *An experimental determination of the excess pressure produced in a liquid dielectric by an electric field*, Proc. Phys. Soc. **80**, 190 (1962).

[29] N. G. C. Astrath, L. C. Malacarne, M. L. Baesso, G. V. B. Lukasievicz, and S. E. Bialkowski, *Unravelling the effects of radiation forces in water*, Nat. Commun. **5**, 4363 (2014).

[30] Z. Li, S. Weilong, P. Nan, and L. Ulf, *Experimental evidence for Abraham pressure of light*, New J. Phys. **17**, 053035 (2015).

[31] R. W. Boyd, *Nonlinear Optics* (Elsevier, 2003).

[32] P. T. Rakich, P. Davids, and Z. Wang, *Tailoring optical forces in waveguides through radiation pressure and electrostrictive forces*, Opt. Express **18**, 14439 (2010).

[33] T. Shimizu, *Two new solutions in cylindrical coordinates of electromagnetic fields of axially traveling waves and some applications*, Electron. Commun. Jpn. Pt. I-Commun. **64**, 37 (1981).

[34] A. W. Snyder and J. D. Love, *Optical Waveguide Theory* (Chapman and Hall, 1983).

[35] D. Marcuse, *Theory of Dielectric Optical Waveguides* (Academic Press, 1974).

[36] D. C. Gazis, *Three-dimensional investigation of the propagation of waves in hollow circular cylinders. I. analytical foundation*, J. Acoust. Soc. Am. **31**, 568 (1959).

[37] J. D. Jackson, *Classical Electrodynamics* (John Wiley & Sons, 2012).

[38] A. S. Biryukov, M. E. Sukharev, and M. D. Evgenii, *Excitation of sound waves upon propagation of laser pulses in optical fibres*, Quantum Electron. **32**, 765 (2002).

[39] M. Aspelmeyer, T. J. Kippenberg, and F. Marquardt, *Cavity optomechanics*, Rev. Mod. Phys. **86**, 1391 (2014).

[40] A. G. Krause, M. Winger, T. D. Blasius, Q. Lin, and O. Painter, *A high-resolution microchip optomechanical accelerometer*, Nat. Photonics **6**, 768 (2012).

[41] C. W. Gardiner and P. Zolle, *Quantum Noise* (Springer, 2004).

[42] M. Mansuripur, *Proc. SPIE* 5930, 59300O (2005).

[43] E. M. Dianov, A. V. Luchnikov, A. N. Pilipetskii, and A. N. Starodumov, *Electrostriction mechanism of soliton interaction in optical fibers*, Opt. Lett. **15**, 314 (1990).

[44] A. Melloni, M. Frasca, A. Garavaglia, A. Tonini, and M. Martinelli, *Direct measurement of electrostriction in optical fibers*, Opt. Lett. **23**, 691 (1998).

[45] G. Agrawal, in *Nonlinear Fiber Optics (Fifth Edition)* (Academic Press, Boston, 2013), pp. 353.

[46] E. Dieulesaint and D. Royer, *Elastic Waves in Solids I: Free and Guided Wave Propagation* (Springer, 2000).

[47] D. K. Biegelsen and J. C. Zesch, *Optical frequency dependence of the photoelastic coefficients of fused silica*, J. Appl. Phys. **47**, 4024 (1976).

[48] W. Primak and D. Post, *Photoelastic constants of vitreous silica and its elastic coefficient of refractive index*, J. Appl. Phys. **30**, 779 (1959).

[49] A. Yariv and P. Yeh, *Optical Waves in Crystals* (Wiley New York, 1984).

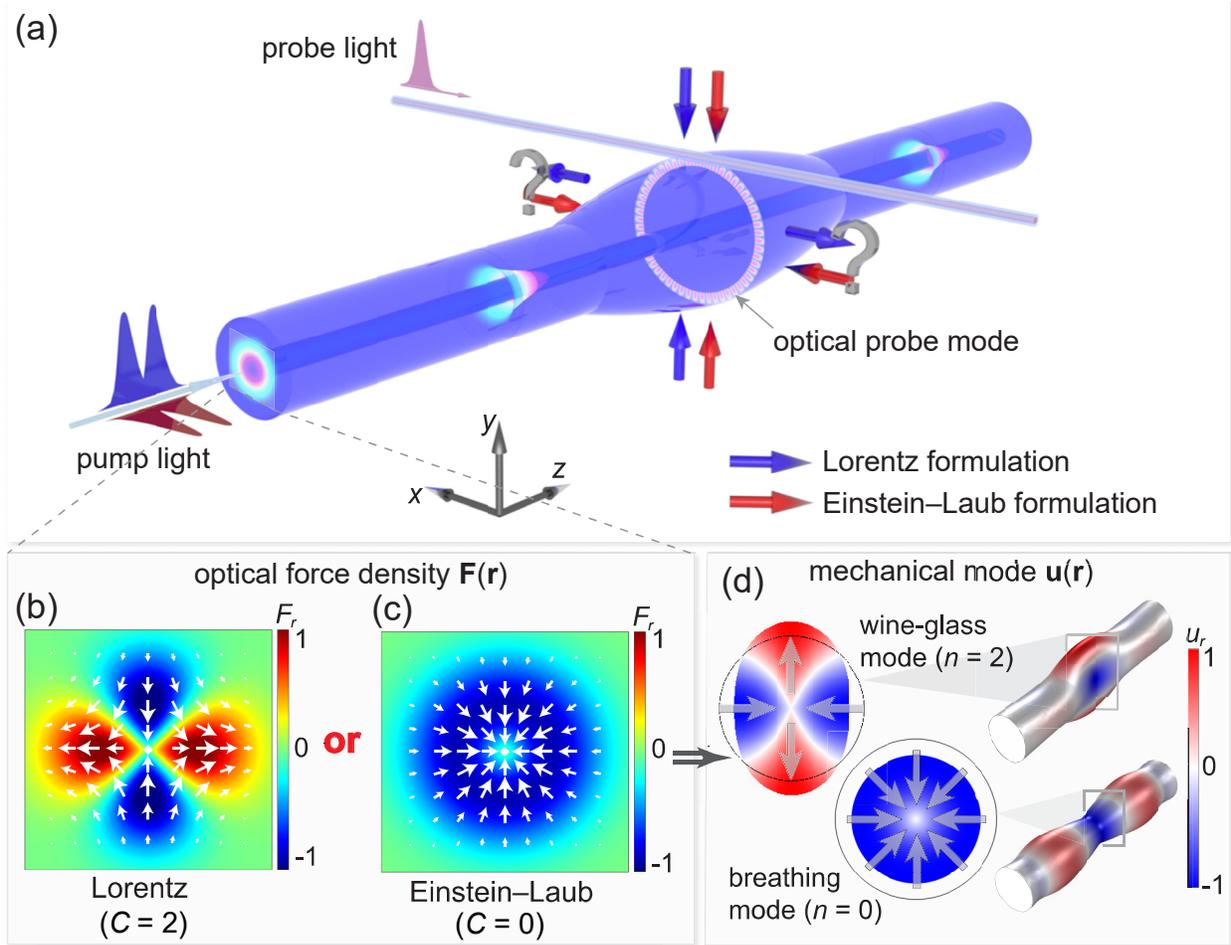

FIG. 1. (a) Schematic of the measurement setup. A linearly polarized Gaussian pump beam is launched into a single-mode fiber, which exerts optical force to actuate the mechanical motion of the fiber. The mechanical motion is detected by a probe field through an ultrahigh-$Q$ optical whispering-gallery mode traveling along the circumference of a fabricated bottle-like microcavity in the fiber. The blue and red arrows denote the force directions inside the optical fiber by a $x$-polarized pump beam in the LO and EL formulations respectively. (b) (c) Calculated force density distributions of the pump beam according to the LO and EL formulations. $F_r$ is the force component in the radial direction, where the outward direction is defined as positive. The LO and EL force have angular symmetry with angular mode number $C = 2$ and $C = 0$, respectively. (d) Profiles of the mechanical wine-glass mode and breathing mode with angular mode number $n = 2$ and $n = 0$, respectively. The arrows indicate the directions of mechanical displacement.

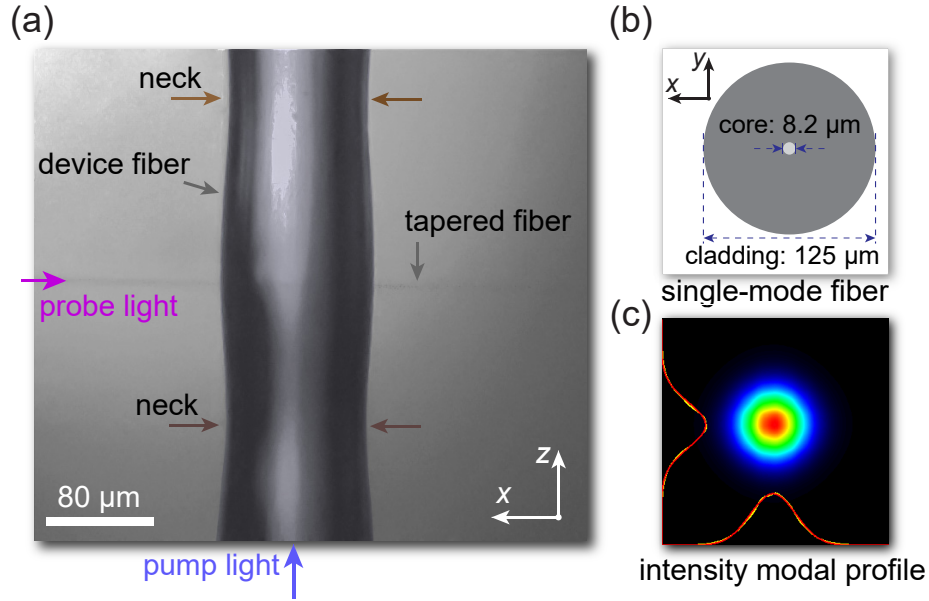

FIG. 2. Device image and experimental configuration. (a) Optical microscope image of the device fiber (showing the part of the bottle-like microstructure) and the tapered fiber in the experimental setup. The pump light beam propagates in the core of the device fiber. The probe light beam is sent through the tapered fiber to be coupled into and out of a whispering-gallery cavity mode supported in the transverse plane of the bottle-like microstructure. (b) Illustration of cross section of the standard single-mode fiber with labeled dimensions. (c) Optical intensity distribution measured from a cut facet of the fiber at the device region, where the yellow and red lines represent the measured and Gaussian-fitted profiles, respectively.

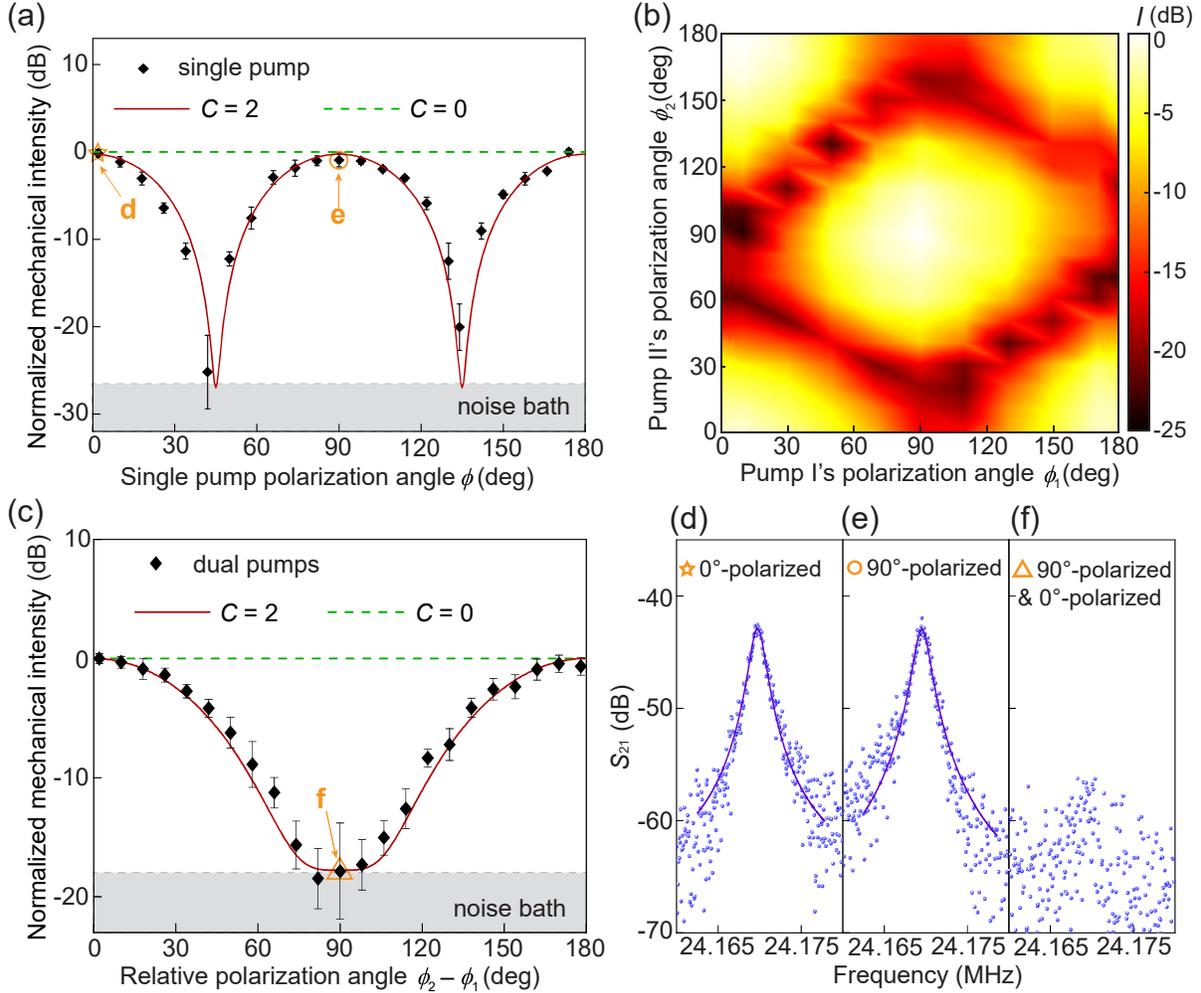

FIG. 3. Dependence of actuated intensity of the wine-glass mode ($n = 2$) on polarization angle(s) of pump beam(s). (a) Measured mechanical intensity as a function of the polarization angle of a single pump beam. (b) Measured mechanical intensity $I$ as a function of the polarization angles of dual pump beams. (c) Measured mechanical intensity as a function of the relative polarization angle $\phi_2 - \phi_1$ of two pump beams, where $\phi_1$ is fixed at 0°. (d), (e), (f) Measured $S_{21}$ spectra showing the relative intensity of the actuated wine-glass mode. The mode in (d) and (e) was actuated by a single pump beam polarized at 0° and 90°, respectively; the mode in (f) was actuated by two pump beams polarized at 0° and 90°. In (a) and (c), the red solid and green dashed lines plot the theoretically predicted results for force with $C = 2$ and $C = 0$, respectively, where the bath noise extracted from the experimental data has been included. The mechanical intensity is normalized to its maximum. The error bars represent one standard deviation from the mean.

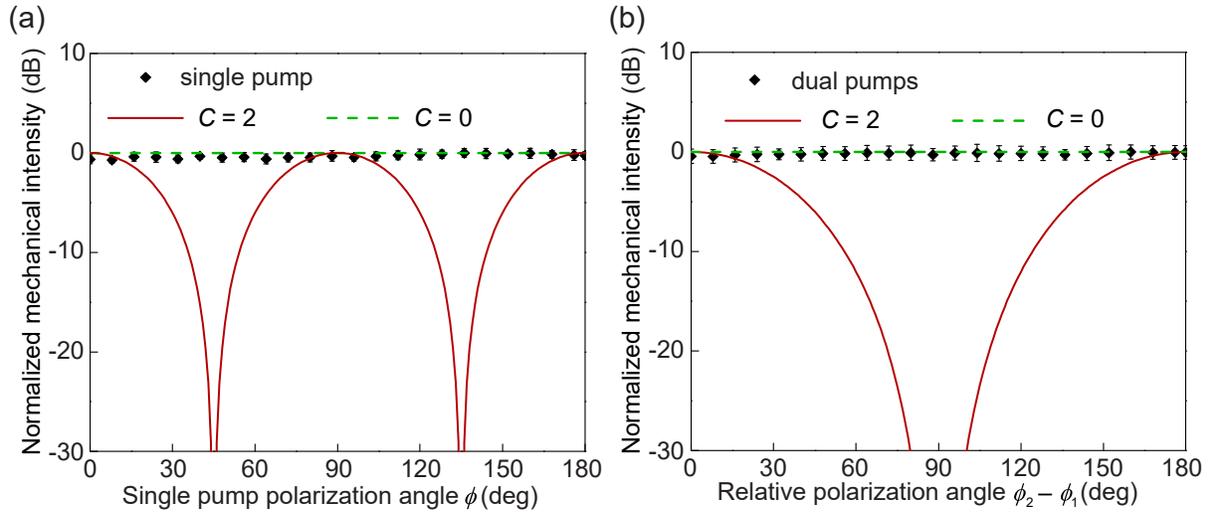

FIG. 4. Dependence of actuated intensity of the breathing mode ($n = 0$) on polarization angle(s) of pump beam(s). (a) Measured mechanical intensity as a function of the polarization angle of a single pump beam. (b) Measured mechanical intensity as a function of the relative polarization angle $\phi_2 - \phi_1$ of dual pump beams, where $\phi_1$ is fixed at 0°. The error bars represent one standard deviation from the mean.

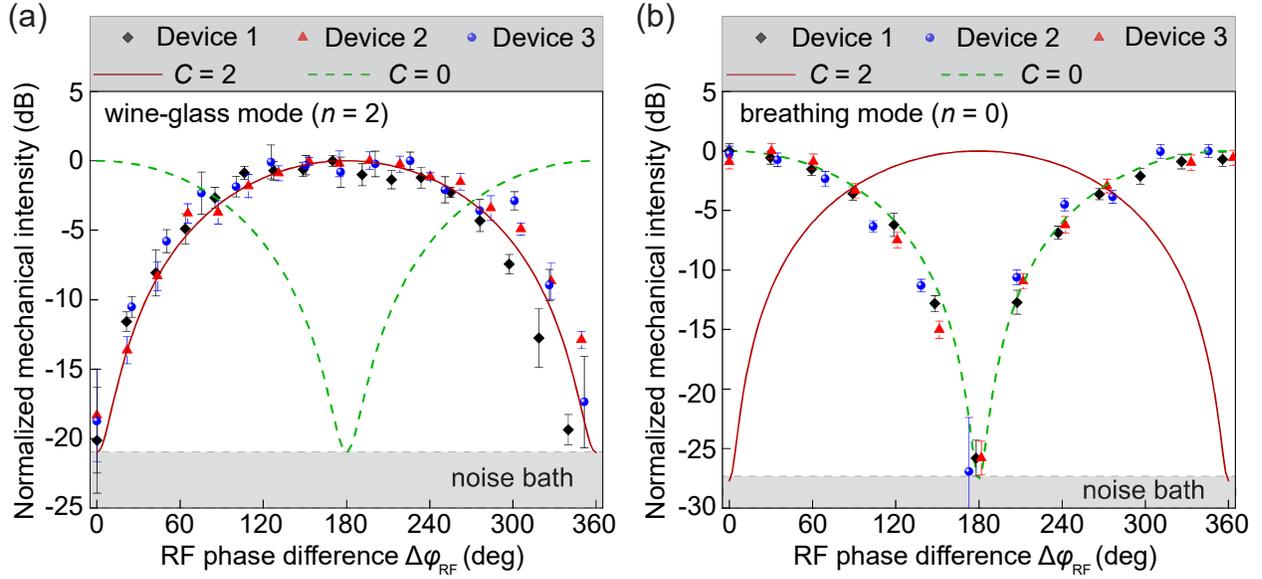

FIG. 5. Dependence of actuated intensity of mechanical modes on the RF modulation phase difference $\Delta\varphi_{RF}$ of two orthogonally polarized pump beams. (a), (b) Measured mechanical intensity of the wine-glass mode ($n = 2$) (a) and the breathing mode ($n = 0$) (b) as a function of the RF modulation phase difference $\Delta\varphi_{RF}$ of the two pump beams. The error bars represent one standard deviation from the mean.